\documentstyle{article}

\begin{document}
\title{3D reduction of the N-body Bethe-Salpeter equation.} 
\author{ J. Bijtebier\thanks{Senior Research Associate at the
 Fund for Scientific Research (Belgium).}\\
 Theoretische Natuurkunde, Vrije Universiteit Brussel,\\
 Pleinlaan 2, B1050 Brussel, Belgium.\\ Email: jbijtebi@vub.ac.be}
\maketitle
\begin{abstract}   \noindent
 We perform a 3D reduction of the two-fermion homogeneous Bethe-Salpeter equation,
by series expansion around a positive-energy instantaneous approximation of the
Bethe-Salpeter kernel, followed by another series expansion at the 3D level in
order to get a manifestly hermitian 3D potential. It turns out that this
potential does not depend on the choice of the starting approximation of the
kernel anymore, and can be written in a very compact form. This result can also
be obtained directly by starting with an approximation of the free propagator,
based on integrals in the relative energies instead of the more usual
$\,\delta-$constraint. Furthermore, the method can be generalized to a system of
N particles, consisting in any combination of bosons and fermions. As an example,
we write the 3D equation for systems of two or three fermions exchanging photons,
in Feynman or Coulomb's gauge.
\end{abstract}
 PACS 11.10.Qr \quad Relativistic wave equations. \newline \noindent PACS
11.10.St \quad Bound and unstable states; Bethe-Salpeter equations.
\newline
\noindent PACS 12.20.Ds \quad Specific calculations and limits of quantum
electrodynamics.\\\\ Keywords: Bethe-Salpeter equations.  Salpeter's equation.
Breit's equation.\par  Relativistic bound states. Relativistic wave equations.
\\\\
 \newpage
\tableofcontents

\section{Introduction.} The homogeneous Bethe-Salpeter equation \cite{1,2} is the
usual tool for computing relativistic bound states. The first difficulty of this
equation comes from the presence of N-1 (for N particles) unphysical degrees of
freedom: the relative time-energy degrees of freedom. In the two-body problem, the
relative energy is usually eliminated by replacing the free Green function by an
expression combining a delta fixing the relative energy and a free 3D
propagator[3-17]. The exact equivalence (in what concerns the physically measurable
quantities of the pure two-fermion problem) with the original homogeneous
Bethe-Salpeter equation can be obtained by recuperating the difference with the
original free Green function in a series of correction terms to the 3D potential.
The dependence in the relative energy of the 4D Bethe-Salpeter equation is thus
transformed into a series of higher-order correction terms to a 3D equation.
This series of 'modified ladder terms" is to be combined with the series of
"crossed terms" which gives the Bethe-Salpeter kernel as a sum of terms
corresponding to irreducible graphs. We do not know if the result converges,
and, if yes, which will be the level of approximation obtained with a given
truncation. This will of course depend on the specific problem studied. The
mutual cancellation between the higher-order "modified ladder terms" (generated
by the 3D reduction) and the "crossed terms", which occurs at the
one-body limit \cite{15,16,17} is an encouraging fact. The starting homogeneous
Bethe-Salpeter equation (for the Bethe-Salpeter amplitude) is deduced from the
inhomogeneous one (for the full propagator) by postulating the presence of a
bound state pole, and is thus a priori valid only for bound states. However, the
scattering amplitude built with the 3D free propagator and potential turns out to
be correct on the mass shell.  
\par
  This kind of 3D reduction
of the two-fermion homogeneous Bethe-Salpeter equation has been performed by many
authors [3-17]. The resulting equations are theoretically equivalent at the limit
of all correction terms included. Of course, in practice, the series generated by
the 3D reduction and the series giving the Bethe-Salpeter kernel itself will both
have to be truncated and the estimation of the level of approximation obtained is
not straightforward.\par 
 A less often used method is based on the replacement of
the Bethe-Salpeter kernel by an "instantaneous" (i.e. independent of the relative
energy) approximation. In this case, the resulting 3D potential is not manifestly
hermitian (for each fixed value of the total energy
on which its depends). In Phillips and Wallace's method, the starting approximation
 is tuned, order by order, is such a way that it becomes the final 3D potential
\cite{19}. In a recent work, we also got an hermitian  3D potential by performing
a  supplementary series expansion
at the 3D level and combining it with the first 3D reducing expansion
\cite{20}.\par 
In the three-fermion problem, a new difficulty comes from the
unconnectedness of the two-body terms of the Bethe-Salpeter kernel, which are in
fact the more important terms and often the only ones to be considered.  In a
constraining propagator-based reduction, the "approached" propagator would have
to be a 3D propagator with two constraints (fixing the two relative energies).
However, when an unconnected two-body term lies between two of these approached
propagators, we get a constraint too much. In Gross' spectator model
\cite{26,27}, the transition matrix element $\,T\,$ is approached by putting on
their positive-energy mass shell the fermion lines which are not going to be
interacting before and after. This means that each three-fermion free propagator
will be "approached" in a way depending on its position in the expansion of
$\,T.$ This leads to a set of Faddeev-like equations, which is not equivalent
to a single equation with a potential, unlike in the nonrelativistic case.   \par 
Recently,
we succeeded in adaptating our two-fermion kernel-based reduction to the
three-fermion case. The 3D potential was a complicated sum of terms of various
origins, although the first-order approximation was still manageable
\cite{20}. When working on an application of this 3D reduction method, we
revisited our two-fermion kernel-based 3D reduction, and found an alternative to
the second series expansion (the one performed at the 3D level to get an hermitian
potential). After combining this series expansion with the first one (given by
the 3D reduction itself), we found that the starting instantaneous approximation
of the Bethe-Salpeter kernel disappears from the final 3D potential.  The result
is in fact a compact expression of the potential that Phillips and Wallace compute
order by order \cite{19}.  Furthermore, it can more directly be obtained by a new
integrating propagator-based reduction method, in which the relative energy is
integrated on, instead of being fixed by a
$\,\delta-$fonction (or constraint). Here again, the physical scattering
amplitude computed at the 3D level is also correct. In fact, the 3D
full propagator turns out to be the retarded part of the 4D full propagator with
the initial and final relative times put to zero, as proposed by Logunov and
Tavkhelidze \cite{4}. \par 
This integrating propagator-based reduction (and the Logunov and
Tavkhelidze's reduction \cite{4,G1,G2}) can easily be adapted
to the three-fermion problem, as the unconnectedness of the two-body parts of the
Bethe-Salpeter kernel is not a difficulty anymore when
integrals on the relative energies are used instead of $\,\delta-$fonctions.
Furthermore, the method can be generalized to a system of N particles,
consisting in any mixing of bosons and fermions. The 3D free propagator is given by
the integral of the product of the N free Green fonctions with respect to the N-1
relative energies and consists in two terms: a term in which all particles have
positive free energies, and a term in which they have all negative free energies
(a slightly different 3D reduction, with simpler first terms, could be obtained by
removing this part of the 3D propagator. We shall consider both options below).
The absence of mixed free-energy signs preserves the 3D equation from continuum
dissolution
\cite{21,22,23,24}. The 3D scattering amplitudes are again correct when all
particles are on their positive-energy mass shell (which has no measurable
consequence when $N\!>\!2\,$), but also in more complicated processes, like the
scattering of a free particle by a bound state.
\par  There are many ways of performing a constraining propagator-based 3D
reduction of the two-body Bethe-Salpeter equation, according to the choice of the
constraint and of the 3D free propagator. Our integrating propagator-based
reduction approach, in contrast, leads to unexpectedly unifying results:
independently of the choice of the initial instantaneous positive-energy kernel,
the final 3D equation is always the same. Furthermore, it can also be obtained by
Phillips and Wallace's method \cite{19} and by the quite different Logunov and
Tavkhelidze's method \cite{4}, while our integrating propagator-based
reduction provides a more compact form for the 3D potential by recombining the
series.  We do not pretend that this 3D reduction converges
more rapidly that the more usual constraining propagator-based reductions (we try
a comparison below). The important result consists in the
fact that it can easily be generalized to more than two particles, giving a 3D
equation for a wave function with a 3D potential, which must necessarily be
truncated but can be indefinitely improved. By comparison, Gross' spectator model
for three particles
\cite{26,27} (which can be considered as a constraining propagator-based
reduction) leads to a set of coupled equations for partial transition operators,
which can not be reduced to a single equation.
\par As a first test of our reduction method, we choose the problems of two and
three fermions in QED, using Feynman's gauge. We compare with a constraining
propagator-based reduction method in the two-fermion case, and with Gross'
spectator model in the three-fermion case. In actual calculations, Coulomb's
gauge would be  a better choice than Feynman's gauge, but it is less suited to a
presentation of our method, as the corrections to the instantaneous approximation
and the three-body terms are then of higher order, competing with a lot of other
contributions.\par
Two natural requirements in the writing of phenomenological equations for systems
of mutually interacting relativistic particles are Lorentz invariance and cluster
separability: if we split the system into clusters by "switching off" selectively
the interactions between the clusters or by by letting the distance between them
become infinite (this is not quite the same thing), we want to get independent sets
of correct covariant equations. It is not easy to satisfy Lorentz invariance and
cluster separability together, and, moreover, to include continuum
dissolution-preserving operators. In our 3D reductions of Bethe-Salpeter
equations the approach is different: the original inhomogeneous Bethe-Salpeter
equation (from which the homogeneous one is derived) is manifestly Lorentz
covariant and cluster separable. The 3D reduction is performed in an unspecified
(at first) reference frame and the individual terms of the series giving the 3D
potential are not covariant, some of them being not cluster separable
either. If we choose to work in the global center of mass frame, all terms can be
made formally covariant, but this reference to the total momentum spoils the
cluster separability. However, if we see the 3D reduction only as a tool to
calculate the bound state spectrum, the exact results can in principle be
indefinitely approached by including more and more higher-order contributions.
Moreover, it turns out that this same 3D equation is still valid for the various
physical scattering amplitudes. The computation of electromagnetic form factors
can also be done by computing electromagnetic scatterings. What is then the
practical meaning of the cluster separability requirement? the selective
"switching off" is clearly a purely mathematical operation which does not
correspond to something physical, while the problem of clusters going to
infinity must be treated as a scattering problem. A practical aspect of
the Lorentz covariance / cluster separability requirement could be the use of, for
example, two-body bound state or scattering data in a three-body context. When
used directly, the two-body data must always be extrapolated off the mass shell,
in a model-dependent way. In our 3D reductions, we could use two-body data
indirectly, through the adjustment of coupling parameters.   
\par\hfill\break In section 2, we
present the usual constraining propagator-based reduction of the two-fermion
Bethe-Salpeter equation, followed by our kernel-based reduction transformed into
an integrating propagator-based reduction, and we compare the first terms of the
3D potentials obtained by both methods. In section 3 we generalize this
integrating propagator-based reduction to N-particle systems (fermions, bosons or
mixings of both). In section 4 we apply our method to systems of two or three
fermions exchanging photons, in Feynman's and Coulomb's gauges.  

\section{The two-fermion problem.}
\subsection{Constraining propagator-based reduction.}
\noindent We shall write the Bethe-Salpeter equation for the bound states
 of two fermions \cite{1} as 
\begin{equation}\Phi = G'^0 K' \Phi,    \label{a}\end{equation}   where
$\Phi$ is the Bethe-Salpeter  amplitude, function of the positions
$x_1,x_2$ or of the momenta 
$p_1,p_2$ of the fermions, according to the representation chosen. The operator
$K'$  is the Bethe-Salpeter kernel, given as a factor of the kernel of an integral
equation in momentum space by the sum of the irreducible two-fermion Feynman
graphs. The operator
$G'^0$ is the product
$G'^0_1G'^0_2$ of the two individual fermion's dressed propagators
\begin{equation}G'^0_i\,=\,{1\over \gamma_i\cdot
p_i\,-\,m_i\,-\,\Sigma_i\,+i\epsilon}\label{b}\end{equation} 
where $\,m_i\,$ is the mass of the fermion $\,i\,$ and $\,\Sigma_i\,(p_i)\,$
the renormalized self-energy function, which becomes proportional to 
$\,(\gamma_i\cdot p_i-m_i)^2\,$ near the mass shell. \par
We shall need a Bethe-Salpeter equation in terms of the free propagator
   
\begin{equation}G^0=G^0_1G^0_2,\qquad G^0_i =
{1\over \gamma_i\cdot
p_i\,-\,m_i\,+i\epsilon}= {1 \over p_{i0}-h_i+i\epsilon
h_i}\,\beta_i 
\label{2}\end{equation}  where the $h_i$ are the Dirac free hamiltonians
\begin{equation}h_i = \vec \alpha_i\, . \vec p_i + \beta_i\, m_i\qquad (i=1,2). 
\label{3}\end{equation}  
In order to transfer the self-energy part of $\,G'^0\,$ to the kernel, we shall
write \cite{25a}  
\begin{equation}K\,=\,K'\,+\,\Sigma\label{c}\end{equation}
with
\begin{equation}\Sigma\,=\,(G^0)^{-1}-(G'^0)^{-1}\,=\,\Sigma_1(G^0_2)^{-1}
+\Sigma_2(G^0_1)^{-1}-\Sigma_1\Sigma_2\label{d}\end{equation}
and get the Bethe-Salpeter equation in the form
\begin{equation}\Phi = G^0 K \Phi.    \label{1}\end{equation}    
We shall define the total (or external, CM, global) and relative (or internal)
variables:
\begin{equation}X = {1 \over 2} (x_1 + x_2)\ , \qquad P = p_1 + p_2\ , 
\label{4}\end{equation} 
\begin{equation}x = x_1 - x_2\ , \qquad p = {1 \over 2} (p_1 - p_2).
\label{5}\end{equation}  and also
\begin{equation}S = h_1 + h_2\ , \quad E = E_1+E_2,\quad E_i=\sqrt{h_i^2}=(\vec
p_i^2+m_i^2)^{1\over 2}.  \label{6}\end{equation} We do not specify the reference
frame in which we write noncovariant quantities like $\,h_i\,$ or $\,E_i.\,$ Our
3D reduction will in fact be frame-dependent. Practically, we shall choose the
global rest frame of the 2- (in this section), 3-, or N-particle system. In the
N-particle problem we shall not try to boost the two-body kernels between the
global rest frame and the (virtual) two-particle rest frames: these relativistic
effects will be taken into account (if desired) by the inclusion of the
higher-order terms of the series generated by the 3D reduction, since this
series, when untruncated, leads to the same measurable quantities as the starting
covariant Bethe-Salpeter equation.\par 
 If we consider the contributions of the poles of 
\begin{equation}G^0={1\over {1\over 2}P_0+p_0-h_1+i\epsilon h_1}
\,{1\over {1\over 2}P_0-p_0-h_2+i\epsilon h_2}\beta_1\beta_2 
\label{15}\end{equation} in an expression like $\,KG^0K,\,$ we must perform an
integration with respect to $\,p_0.\,$ If $\,K\,$ is instantaneous, we get
\begin{equation}\int dp_0 G^0(p_0)\,=\,-2i \pi\tau\, g^0\,  
\beta_1 \beta_2,\qquad g^0\,=\,{1 \over P_0-S+i\epsilon P_0 } 
\label{18}\end{equation} where
\begin{equation}\tau={1\over 2} (\tau_1 + \tau_2), \qquad \tau_i = {h_i
\over \sqrt{h_i^2}} = {h_i \over E_i}
\label{19}\end{equation}  can also be written
\begin{equation}\tau=\Lambda^{++}-\Lambda^{--}, \qquad
\Lambda^{ij}=\Lambda_1^i\Lambda_2^j,
\qquad \Lambda_i^\pm={E_i\pm h_i\over 2E_i}. \label{20}\end{equation}      When
$\,K\,$ is not instantaneous, we must add the contributions of its singularities.
Furthermore, in the residues of the poles of
$\,G^0\,$ we must take $\,K\,$ at $\,p_{10}\!=\!h_1\,$ (i.e.  at $\,+E_1\,$ in the
subspace built with the $\,h_1\!=\!+E_1\,$ eigenstates of $\,h_1,\,$ and at
$\,-E_1\,$ in the complementary subspace)     or at
$\,p_{20}\!=\!h_2,\,$ according to the chosen integration path and to the sign of
$\,\tau.\,$ We shall perform a 3D reduction based on the replacement of the free
propagator $\,G^0\,$ by a carefully chosen expression  
\begin{equation}G^{\delta}(p_0)=-2i \pi\,\tau\,\delta(p_{10}\! -\!h_1)\,g^0\,  
\beta_1 \beta_2\label{19a}\end{equation} combining the 3D propagator $\,g^0\,$
with the constraint
\begin{equation}\delta(p_{10}\! -\!h_1)\equiv\Lambda_1^+\delta(p_{10}\!
-\!E_1)+\Lambda_1^-\delta(p_{10}\! +\!E_1)=\delta(p_0\! -\!s_1),\qquad
s_1={P_0\over2}+h_1.\label{19b}\end{equation}  
 The operator $\tau$ has a clear meaning in the basis built with the free
solutions: it is +1 for
$h_1,h_2>0$, -1 for
$h_1,h_2<0$ and zero when they have opposite signs. It comes from the dependence
of the $p_0$ integral on the signs of the $i\epsilon h_i$. \par This choice of
$\,G^{\delta}\,$ has three merits: a) It leads directly to a simple
equation (Salpeter's equation without higher-order correction terms to the
potential) with an instantaneous approximation of the kernel. The higher-order
correction terms (necessary to get the correct one-body limits) will then be the
consequence of the dependence of the kernel in the relative energy. b) In the
two-fermion plus potential problem and in the three-fermion problem, the operator
$\,\tau\,$ prevents the mixing of asymptotically free fermions with opposite
energy signs, which is the origin of the continuum dissolution problem
\cite{21,22,23,24}  c) It preserves a particle-antiparticle symmetry, which is a
characteristic feature of relativistic theories. We could also replace
$\,\tau\,$ by $\,\Lambda^{++},$ as the $\,\Lambda^{--}\,$ part   does not
contribute much in practice. One of the fermions would then be on its
positive-energy mass shell, as in Gross' spectator model \cite{16}. There is an
infinity of other possible choices, like fixing $\,p_0\,$ at its on mass shell
value
$(h_1\!-\!h_2)/2\,$, instead of fixing $\,p_{10}\,$ at its on mass shell value
$\,h_1\,$, treating thus both fermions in a symmetrical way. With Sazdjian's
propagator \cite{18}, one gets an explicitly covariant equation (we
recently combined this propagator with a kind of covariant positive-energy
projector \cite{31}). With Lepage \cite{9}, we get directly a 3D
equation written in Schr\"odinger's form. An important feature is the presence
or absence of a continuum dissolution-preserving operator like $\,\tau\,$ or
$\,\Lambda^{++}.$ Without such an operator, the propagator could not be used
beyond the two-body problem, including the two-body in an external potential
problem. Our constraining operator contains thus the operator $\,\tau\,$ (we
shall also examine the $\,\Lambda^{++}\,$ choice). Besides that, the
principal reason of our choice (\ref{19a}) is to allow for an easy term by
term comparison with our integrating propagator-based reduction below.       
\par    Let us now write the free propagator as the sum of the zero-order
propagator, plus a remainder:
\begin{equation}G^0=  G^{\delta}+G^R.  \label{23}\end{equation}  The
Bethe-Salpeter equation  becomes then the inhomogeneous equation
\begin{equation}\Phi=G^0K\Phi=(G^\delta +G^R)K\Phi=\Psi +G^RK\Phi,
\label{24}\end{equation}  with
\begin{equation}\Psi=G^\delta K\Phi \qquad (=G^\delta (G^0)^{-1}\Phi).
\label{25}\end{equation}  Solving (formally) the inhomogeneous equation
(\ref{24})  with respect to $\,\Phi\,$ and putting the result into (\ref{25}), we
get
\begin{equation}\Psi=G^\delta K(1-G^RK)^{-1}\Psi=G^\delta K^T\Psi  
\label{26}\end{equation}  where
\begin{equation}K^T=K(1-G^RK)^{-1}=K+KG^RK+...=(1-KG^R)^{-1}K 
\label{27}\end{equation}  obeys
\begin{equation}K^T=K+KG^RK^T=K+K^TG^RK. \label{28}\end{equation}  The reduction
series (\ref{27}) re-introduces in fact the reducible graphs into the
Bethe-Salpeter kernel, but with $G^0$ replaced by $G^R$. Equation (\ref{26}) is a
3D equivalent of the Bethe-Salpeter equation. \par
 The relative energy dependence of eq. (\ref{26}) can be easily eliminated:
\begin{equation}\Psi=\delta(p_0\! -\! s_1)\,\psi \label{29}\end{equation}  and
$\,\psi\,$ obeys:
\begin{equation}\psi\,=-2i\pi\,g^0\, \tau \,\int dp_0' dp_0
\delta(p'_0\! -\!s_1)\beta_1\beta_2K^T(p_0',p_0)\delta(p_0\!
-\!s_1)\,\psi.\label{32a}\end{equation} Using the identity $\,\psi\!=\! \tau^2
\psi,\,$ we can write 
\begin{equation}\psi\,=g^0\,\tau\,V\,\psi,\qquad
V\,=-2i\pi\,\tau^2\,\beta_1\beta_2K^{T}(s_1,s_1)\,\tau^2,\label{30}\end{equation} 
\begin{equation}\beta_1\beta_2K^{T}(s_1,s_1)\,\equiv \!\int dp_0'
dp_0\delta(p'_0\! -\!s_1)\beta_1\beta_2K^T(p_0',p_0)\delta(p_0\!
-\!s_1).\label{32}\end{equation} Note that we write
$\,(p'_0,p_0)\,$ but
$\,(s_1,s_1),\,$ as we keep $\,s_1\,$ in operator form. This operator can be
diagonalized in the spatial momentum space by using the
$\,\Lambda^{ij}\,$ projectors. The eigenvalue will depend on the position of
$\,s_1\,$ in the formula: the eigenvalue of the first $\,s_1\,$ in (\ref{32})
will be built with the final momenta and that of the last
$\,s_1\,$ will be built with the initial momenta.  \par The inversion of the
reduction is given by
\begin{equation}\Phi\,=\,(\,1-G^RK\,)^{-1}\,\Psi\,=\,
(\,1+G^RK^T\,)\,\Psi\,=\,(\,1+G^0K^T-G^{\,\delta}
K^T\,)\,\Psi\,=\,G^0K^T\,\Psi\label{32a*}\end{equation} or, explicitating the
relative energy:
\begin{equation}\Phi(p'_0)\,=\,G^0(p'_0)\,K^T(p'_0,s_1)\,\psi.
\label{32b}\end{equation}   
\par    The splitting of $\,G^0\,$ into two terms containing a
$\,\delta\,$ is the origin of unphysical singularities in the terms of
$\,K^T\,$ when the argument of the delta vanishes on the singularities of
$\,K.\,$ When the full $\,K^T\,$ is computed, the singularities of the different
terms cancel mutually. When $\,K^T\,$ is truncated, some of the unphysical
singularities have to be removed by hand \cite{16,26}.\par This 3D reduction can
also be described in terms of transition operators. The 4D transition operator 
corresponding to our modified Bethe-Salpeter equation (\ref{1}) is
\begin{equation}T\,=\,K\,+\,K\,G^0\,K\,+\,\cdots\label{*21}\end{equation}
\noindent and $\,K^T\,$ can be obtained by keeping only the
$\,G^R\,$ part of $\,G^0\,$ in it. We have also
$$T=K(1\!-\!G^0K)^{-1}=K(1\!-\!G^RK\!-\!G^{\delta}K)^{-1}=K(1\!-\!G^RK)^{-1}
(1\!-\!G^{\delta}K(1\!-\!G^RK)^{-1})^{-1}$$
\begin{equation}=K^T(1\!-\!G^{\delta}K^T)^{-1}=K^T+K^TG^{\delta}K^T+\cdots
\label{*22}\end{equation} 
 so that the 3D transition operator 
\begin{equation}T^{3D}\,=\,V\,+\,V\,g^0\,\tau\,V\,+\,\cdots
\label{*23}\end{equation}
is also given by
\begin{equation}T^{3D}\,=\,-2i\pi\,\tau^2\,\beta_1\beta_2\,T(s_1,s_1)\,
\tau^2.\label{*24}\end{equation}
The operator $\,T\,$ and the operator $\,T'\,$  corresponding to the original
Bethe-Salpeter equation (\ref{a}) are related by the common full propagator
$\,G:$ 
\begin{equation}G^0+G^0TG^0\,=\,G\,=\,G'^0+G'^0T'G'^0\label{a1}\end{equation}
with
\begin{equation}G'^0\,=\,G^0\,(1-\Sigma\,G^0\,)^{-1}\,=\,
G^0_1\,(1-\Sigma_1\,G^0_1\,)^{-1}G^0_2\,(1-\Sigma_2\,G^0_2\,)^{-1}.
\label{a2}\end{equation}
 We see that the 3D transition operator is a constrained form of that of field
theory.  When
both fermions are on their positive-energy mass shells, the operators
$\,T(s_1,s_1), T,T',\,$ become equal and $\,T^{3D}\,$ becomes proportional to the
physical scattering amplitude.
 This was not guaranteed a priori, as our original two-fermion Bethe-Salpeter
equation (\ref{1})  was valid only for bound states. Our 3D equation (\ref{30})
is a bound state equation too. To include the scattering states we should add an
inhomogeneous term, or write the equation in the form
\begin{equation}(P_0-S\,)\,\psi\,=\tau\,V\,\psi.\label{38.2}\end{equation}

\subsection{Kernel-based reduction.} 
A few operators in this kernel-based reduction will be given the same name as
similar operators used in the constraining propagator-based reduction, although
being not identical to them. In order to avoid any confusion, we shall overwrite
these operators with a $\,\,\,\widetilde {}\,\,$ .\par
 If the Bethe-Salpeter kernel were
instantaneous (i.e. independent of
$\,p_0\,$), one would get Salpeter's equation by integration with respect to
$\,p_0\,$ \cite{3}. In the realistic case of a non-instantaneous kernel, it is
possible to compute the bound state energies at the 4D level by perturbations
around an instantaneous approximation of the kernel \cite{11,IZ,15}. Here, we
want to build a 3D reduction around an approximation
$\,K^0\,$ of the Bethe-Salpeter kernel. Let us write
\begin{equation}K\,=\,K^0\,+\,K^R.\label{38.1}\end{equation} The Bethe-Salpeter
equation becomes
\begin{equation}\Phi\,=\,G^0K^0\Phi\,+\,G^0K^R\Phi\label{39}\end{equation}
\begin{equation}\Phi\,=\,(1-G^0K^R\,)^{-1}G^0K^0\Phi\,=\,G^KK^0\Phi
\label{40}\end{equation}  with
\begin{equation}G^K\,=\,G^0+G^0K^RG^0+\cdots\equiv
G^0+G^{KR}.\label{41}\end{equation} If we now specialize $\,K^0\,$ to an
instantaneous positive-energy kernel
\,($K^0\!=\!\Lambda^{++}\beta_1\beta_2K^0\Lambda^{++}\,$ and is independent of
$\,p_0),$ we get
\begin{equation}\phi\,=\,(\,g^0+g^{KR}\,)\,V^0\,\phi\label{42}\end{equation} with
\begin{equation}\phi\,=\,\Lambda^{++}\int dp_0\,\Phi(p_0)\,,\qquad
V^0\,=\,-2i\pi\,\beta_1\beta_2\,K^0\,,\label{43}\end{equation}
\begin{equation}g^{KR}\,=\,{-1\over2i\pi}\,\Lambda^{++}\int
dp'_0dp_0\,G^{KR}(p'_0,p_0)\,\beta_1\beta_2\label{44}\,\Lambda^{++}.\end{equation}
The interaction term $\,(g^0)^{-1}(\,g^0+g^{KR}\,)\,V^0\,$ is not hermitian (for
a fixed value of the total energy on which its depends), so that we do not know if
equation (\ref{42}) will have a real energy spectrum. Let us write this equation
in the form
\begin{equation}\phi=(1+g^0T^{KR})\,g^0V^0\phi\,,\qquad
T^{KR}=(g^0)^{-1}g^{KR}(g^0)^{-1}\label{46}\end{equation} in order to perform the
transformations
\begin{equation}(1+g^0T^{KR})^{-1}\phi\,=\,g^0V^0\phi\label{47}\end{equation}
$$\phi\,=\,\big[\,g^0V^0+1-(1+g^0T^{KR})^{-1}\,\big]\,\phi
=\,g^0\,\big[\,V^0+T^{KR}(1+g^0T^{KR})^{-1}\,\big]\,\phi$$
\begin{equation}=g^0\big[<K^0>+<K^R(1-G^0K^R)^{-1}>
(1+g^0<K^R(1-G^0K^R)^{-1}>)^{-1}\,\big]\,\phi\label{48}\end{equation} with the
definition
\begin{equation}<A>={-1\over2i\pi}\Lambda^{++}(g^0)^{-1}\int
dp'_0dp_0\,G^0(p'_0)A(p'_0,p_0)G^0(p_0)\,\beta_1\beta_2\,\Lambda^{++}(g^0)^{-1}
.\label{49}\end{equation} Let us expand $\,g^0\,$ into a 4D operator:
\begin{equation} G^I=\,\,\,\,\,>g^0\!<\label{50}\end{equation} 
defined as (writing the relative energy arguments, the other momentum arguments
$\,P\,$ and $\,\vec p\,$ remaining local):
\begin{equation} G^I(p'_0,p_0)\,=\,G^0(p'_0)\,\beta_1\beta_2\,{\Lambda^{++}
\over-2i\pi\,g^0}\,G^0(p_0)\label{51}\end{equation} and write (\ref{48}) in the
form
$$\phi\,=\,g^0\,<K^0+K^R(1-G^0K^R)^{-1} (1+ G^IK^R(1-G^0K^R)^{-1})^{-1}>\,\phi$$
\begin{equation}=\,g^0\,<K^0+K^R (1-G^0K^R+ G^IK^R)^{-1}>\,\phi
=\,g^0\,<K^0+K^R (1-\widetilde G^{R}K^R)^{-1}>\,\phi\label{52}\end{equation} with
\begin{equation}\widetilde
G^{R}(p'_0,p_0)\,=\,G^0(p_0)\,\delta(p'_0-p_0)-\, G^I(p'_0,p_0).\label{53}\end{equation} It is easy to see that
$\,A\widetilde G^{R}\!=\!\widetilde G^{R}A\!=\!0\,$ whenever $\,A\,$ is an
instantaneous positive-energy operator like $\,K^0.\,$ We can then recombine
$\,K^0\!+\!K^R\!=\!K\,$ in the 3D equation, which becomes
\begin{equation}\phi =\,g^0\,\widetilde V\,\phi,\qquad\widetilde V
=\,<\widetilde K^T>,\,\qquad \widetilde K^T\,=\, K (1-\widetilde
G^{R}K)^{-1}.\label{54}\end{equation}
Computing the scattering matrix element as in the preceding subsection gives
\begin{equation}\widetilde
T^{3D}\,\equiv\,\,\widetilde V(1-g^0\widetilde
V)^{-1}\,=\,\,<T>.\label{55}\end{equation} 
On the positive-energy mass shells $\,P_0\!=\!E'_1\!+\!E'_2\!=\!E_1\!+\!E_2\,$ the
operators $\,(g^0)^{-1}\,$  of  (\ref{49}) vanish and we remain with the
residues of the corresponding poles in the integrations with respect to
$\,p'_0\,$ and $\,p_0,\,$ so that
\begin{equation}<T>\,=\,-2i\pi\,\Lambda^{++}\beta_1\beta_2\,T\,(\,p'_{i0}=E_i,\,
p_{i0}=E_i\,)\,\Lambda^{++}\label{55c}\end{equation}
and we have again the correct physical scattering amplitude ($\,T\,$ is
written here with operators as arguments, as in (\ref{32})). 
\par 
Equation (\ref{55}) can be inverted to give $\,\widetilde V\,$ in terms of $\,<T>:$
\begin{equation}\widetilde V\,=\,<T>(\,1+g^0<T>)^{-1}\,=\,<\widetilde
K^T>.\label{55bis}\end{equation}   
In fact, it is possible to use equation (\ref{55}) to define the 3D reduction.
The corresponding 3D full propagator
\begin{equation}g\,=\,g^0\Lambda^{++}\,+\,g^0\,\widetilde
T^{3D}g^0\label{55ter}
\end{equation}
is then proportional to the integral of $\,G\,$ with respect to the initial and
final relative energies, between two $\,\Lambda^{++}\,$ projectors. This
manipulation preserves the position of the bound state poles and the physical
scattering amplitudes. In configuration space, it means that we take the
retarded part of $\,G\,$ at equal times. The transformation of the
inhomogeneous Bethe-Salpeter equation into an equation for $\,g\,$ was
the starting point of the 3D reduction of Logunov and Tavkhelidze
\cite{4}.         
\par  Phillips and Wallace's
3D reduction is a kernel-based reduction in the
$\,\tau^2\!=\!1\,$ subspace. Their way of symmetrizing the interaction term
consists in tuning $\,K^0\,$ in order to make $\,g^{KR}\,$ vanish, considering
thus
$\,g^{KR}\!=\!0\,$ as an equation in $\,K^0\,$ to be solved order by order. If we
do that here in the $\,\Lambda^{++}\!=\!1\,$ subspace, we see that it implies
that the operator at the left of (\ref{47}) is the identical operator. The
comparison with (\ref{54})   leads to
\begin{equation}-2i\pi\,\beta_1\beta_2\,K^0\,=\, <K(1-\widetilde
G^{R}K)^{-1}>.\label{55b}\end{equation}
Our 3D equation (\ref{54}) is quite general, as it does not depend on the
initial choice of $\,K^0\,$ anymore (at the 3D level, it is of course still
possible to perform a perturbation calculation starting with $\,<K^0>,\,$
using (\ref{52})).  This 3D reduction is also identical to that of Phillips and
Wallace \cite{19}, obtained by tuning an unspecified instantaneous kernel,
and to that of Logunov and Tavkhelidze \cite{4}, obtained by taking the
retarded part of the full propagator $\,G\,$ at equal times. Our own 3D
reduction gives a compact expression of the 3D potential, by recombining the
series.\par
The independence of our 3D equation on the choice of $\,K^0\,$ suggests that
it could also be obtained by a new kind of propagator-based reduction, in
which the relative energies are integrated on instead of being "constrained".
We shall now present this new approach, and we shall see in section 3
that is can be easily generalized to N-body systems.    
\par     

\subsection{Integrating propagator-based reduction.} We shall now perform directly
an integrating propagator-based reduction, inspired by our kernel-based reduction
above, but we shall work in the larger
$\,\tau^2\!=\!1\,$ subspace, defining
\begin{equation} G^I(p'_0,p_0)\,=\,G^0(p'_0)\,\beta_1\beta_2\,{\tau
\over-2i\pi\,g^0}\,G^0(p_0)\label{55c}\end{equation}
\begin{equation}G^0=   G^I+\widetilde G^R. 
\label{56}\end{equation}  
\begin{equation}\Phi=G^0K\Phi=( G^I +\widetilde
G^R)K\Phi=\widetilde \Psi +\widetilde G^RK\Phi,
\label{57}\end{equation}  with
\begin{equation}\widetilde \Psi= G^I K\Phi= G^I
K(1-\widetilde G^RK)^{-1}\,\widetilde \Psi= G^I \widetilde
K^T\widetilde \Psi  
\label{58}\end{equation}  where
\begin{equation}\widetilde K^T=K(1-\widetilde G^RK)^{-1}=K+K\widetilde
G^RK+...=(1-K\widetilde G^R)^{-1}K. 
\label{59}\end{equation}   When we explicitate the relative energies in equation
(\ref{58}), we get
$$\widetilde \Psi(p''_0)\,=\,G^0(p''_0)\,\beta_1\beta_2\,{\tau
\over-2i\pi\,g^0}\int dp'_0dp_0\,G^0(p'_0)K(p'_0,p_0)\,\Phi(p_0)$$
$$=\,G^0(p''_0)\,\beta_1\beta_2\,{\tau
\over-2i\pi\,g^0}\,\int dp'_0\,\Phi(p'_0)$$
\begin{equation}=\,G^0(p''_0)\,\beta_1\beta_2\,{\tau
\over-2i\pi\,g^0}\int dp'_0dp_0\,G^0(p'_0)\widetilde
K^T(p'_0,p_0)\,\widetilde \Psi(p_0).\label{60}\end{equation} Defining now
\begin{equation}\phi\,=\,\tau\int dp_0\,\Phi(p_0)\label{61}\end{equation}
\begin{equation}<A>\,=\,{-1\over2i\pi}\,\tau^2(g^0)^{-1}\int
dp'_0dp_0\,G^0(p'_0)A(p'_0,p_0)G^0(p_0)\,\beta_1\beta_2\,\tau^2\,(g^0)^{-1}
\label{62}\end{equation} so that
\begin{equation} G^I=\,\,\,\,>\tau
g^0\!<\label{63}\end{equation}  and writing (\ref{60}) in terms of $\,\phi,\,$ we
get
\begin{equation}\phi\,=\,g^0\,\tau\,<\widetilde
K^T>\,\phi.\label{64}\end{equation}     If we perform again our kernel-based
reduction of subsection 2.2, but now in the $\,\tau^2\!=\!1\,$ subspace and using
the definitions (\ref{61}-\ref{63}), we get
\begin{equation}\chi\,=\,g^0\,\tau\,\big[\,<K^0>-\,\tau<K^0>\tau\,+\,
\tau<\widetilde
K^T>\tau\,\big]\,\chi\,,\qquad\chi=\tau\phi\label{65}\end{equation} and $\,K^0\,$
will disappear if it does not connect the (++) and the (-\,-) subspaces, leading
to our equation (\ref{64}). Phillips and Wallace's potential, which must reappear
after the symmetrizing transformation, will be $\,<K^0>=<\widetilde K^T>.$
\par
If we prefer to work in the $\,\Lambda^{++}\!=\!1\,$ subspace, we can
recover the results of subsection 2.2 by simply replacing $\,\tau\,$ by
$\,\Lambda^{++}.$         
        
\subsection{Comparison of the lowest-order terms.} With the integrating propagator-based reduction, the final 3D potential will be
$$\widetilde V\,=\,\,\,<K>+<K\,\widetilde G^RK>+\cdots$$
\begin{equation}=\,\,\,<K>+<K\, G^0K>-<K>\tau
g^0\!<K>+\cdots\label{66}\end{equation} At order 4 in the coupling constant, we
should keep the ladder term and the first crossed term in $\,<K>\,$ and only the
ladder term in $\,<K\,\widetilde G^RK>.\,$ The contributions of the higher-order
terms should a priori decrease with the number of $\,\widetilde G^R:\,$ once the
leading contributions containing the poles in
$\,P_0\!-\!(E_1\!+\!E_2)\,$ have been removed, we remain with the smaller
contributions coming from the residues of the poles of $\,K\,$ or from its
(+\,--) and (--\,+) components.\par For comparison with the usual constraining propagator-based reduction of subsection 2.1, we shall split $\,G^0\,$ into
$\,G^{\delta}\!+\!G^R\,$ in the definition (\ref{49}) of
$\,<\!A>\,$ (it must be noted that this splitting introduces unphysical
singularities \cite{16,26}). Keeping only the terms with zero or one
$\,G^R,\,$ we get
$$\widetilde V\approx\,<K>^{\delta\delta}+<KG^RK>^{\delta\delta}$$
\begin{equation}
+\,\,\big[1-<K>^{\delta\delta}\tau g^0\big]<K>^{R\delta}+<K>^{\delta R}\big[1-\tau
g^0<K>^{\delta\delta}\big]\label{67}\end{equation} where the index $\,R\,$ or
$\,\delta\,$ indicates which part of $\,G^0\,$ has been kept. The two first terms
are also the two first terms of the constraining propagator-based reduction. The
two other terms would not contribute at order 4 in a perturbation calculation
beginning with
$\,<K>^{\delta\delta}.\,$ We see the kind of rearrangements by which the
constraining propagator-based reduction and the integrating propagator-based
reduction will finally lead to the same energy spectrum. At order 2 only in the
coupling constant, we would get
\begin{equation}\widetilde V\approx\,<K>^{\delta\delta}+<K>^{R\delta}+<K>^{\delta
R}\label{68}\end{equation} and checking if the addition of the two last terms
brings (\ref{68}) nearer to (\ref{67}) could help to choose between both
approaches in a specific problem. \par
The 3D free propagator is $\,\tau g^0\,$ in the constraining propagator as in
the integrating propagator-based reductions. The 3D potentials
$\,V\,$ and
$\,\widetilde V\,$ are both hermitian and both equations are equivalent. We
must however not conclude that these potentials are equal: the dependence of
the potential on the total energy allows for an infinity of equivalent
equations.     
      
\section{N-body problem.}
\subsection{Generalization of the integrating propagator-based reduction.} The
integrating propagator-based reduction is a good candidate to a double
generalization: from two to N fermions and/or from N fermions to any system of
f\,=\,0,1,...N fermions with b\,=\,N-f bosons. The propagators of the fermion i
and the boson j are respectively
\begin{equation}G^0_i\,=\,{1\over p_{i0}-h_i+i\epsilon
h_i}\,\beta_i\,,\label{69}\end{equation}
\begin{equation}G^0_j\,=\,{1\over
p^2_{j0}-E^2_j+i\epsilon}\,=\,{1\over2E_j}\,\sum_{\sigma_j}\, {\sigma_j\over
p_{j0}-\sigma_jE_j+i\epsilon\sigma_j}\label{69b}\end{equation} with
$\,\sigma_j\!=\!\pm1.\,$ The free propagator $\,G^0\,$ for a system of
$\,f\!=\!0,1...N\,$  fermions and b\,=\,N-f bosons will be
\begin{equation}G^0\,=\,G^0_1\,...\,G^0_f\,G^0_{f+1}\,...\,G^0_N
\label{70}\end{equation}      and the corresponding 3D propagator will be
proportional to
\begin{equation} \int
dp_0\,G^0(p_0)\,\equiv\int\delta\,(P_0-\sum_{i=1}^Np_{i0})\,dp_{10}...
dp_{N0}\,G^0(p_{10},...p_{N0}).\label{71}\end{equation} We shall perform the
integral in $\,p_{10}\,$ by replacing it by
$\,(P_0\!-\!\sum_{i=2}^N\,p_{i0})\,$ in the first propagator. Each other
$\,p_{i0}\,$ will then appear twice: in $\,G^0_1\,$  and in the corresponding
$\,G^0_i,\,$ and the integral with respect to $\,p_{i0}\,$ will be zero unless
$\,h_i\,$ (or $\,\sigma_i\,$) and $\,h_1\,$ (or
$\,\sigma_1\,$) have the same sign. The final result will be
\begin{equation}\int
dp_0\,G^0(p_0)\,=\,{(-2i\pi)^{N-1}\over\omega}\,\tau\,g^0\,\beta
\label{71a}\end{equation} with
\begin{equation}g^0\,=\,{1\over P_0-S+i\epsilon P_0}\,,\qquad
S=E\,(\Lambda^+-\Lambda^-)\,,\qquad
\beta=\beta_1...\beta_f\,,\label{72}\end{equation}
\begin{equation}\Lambda^{\pm}\,=\Lambda_1^{\pm}\,...\,\Lambda_f^{\pm}
\,,\qquad\tau=\Lambda^++(-)^{f+1}\Lambda^-\,,\label{73}\end{equation}
\begin{equation}E\,=\,\sum_{i=1}^N\,E_i\,,\qquad\omega\,=\,2^b\,E_{f+1}
\,...\,E_{N}\label{74}\end{equation} for $\,f\neq0,\,$ so that
\begin{equation}\tau\,g^0\,=\,{\Lambda^+\over
P_0-E+i\epsilon}\,+\,(-)^{f+1}\,{\Lambda^-\over
P_0+E-i\epsilon}.\label{75}\end{equation} When $\,f\!=\!0,\,$ (bosons only), we
have no $\,\tau\,$ and no $\,\beta\,$ (we can replace them by 1 in (\ref{71a}))
and
\begin{equation}g^0\,=\,{1\over P_0-E+i\epsilon}\,-\,{1\over
P_0+E-i\epsilon}\,=\,{2E\over
P_0^2-E^2+i\epsilon}\,.\label{76}\end{equation}                  The
Bethe-Salpeter equation for N particles can still be written
\begin{equation}\Phi\,=\,G^0\,K\,\Phi\label{77}\end{equation}
with
\begin{equation}K\,=\,K'\,+\,\Sigma,\qquad\Sigma\,=\,(G^0)^{-1}-(G'^0)^{-1}
\label{dd}\end{equation}
 where $\,K'\,$ will
 be given by a combination of irreducible $\,2\le n\le N-$body irreducible
kernels. For N=3:
\begin{equation}K'\,=\,K'_{12}\,(G'^0_3)^{-1}\,+\,
K'_{23}\,(G'^0_1)^{-1}\,+\,K'_{31}\,(G'^0_2)^{-1}\,+\,
K'_{123}.\label{78}\end{equation}
For  $\,N\ge4\,$ the writing of $\,K'\,$ becomes more complicated, because of
the presence of commutating kernels like $\,K'_{12}\,$ and $\,K'_{34}\,$ which
would lead, without corrections, to an overcounting of some graphs in the
expansion of $\,G.\,$ We examine this problem elsewhere \cite{29}.
\par      Let us now define a N-body operator
$\, G^I:$
\begin{equation} G^I(p'_0,p_0)\,=\,G^0(p'_0)\,\beta\,{\tau\,\omega
\over(-2i\pi)^{N-1}\,g^0}\,G^0(p_0)\label{79}\end{equation} which is such that
$$\int dp_0\, G^I\,(p'_0,p_0)\,=\,G^0(p'_0)\,\beta\,\tau^2\beta\,=\,\,\tau^2\,G^0(p'_0),$$
\begin{equation}
\qquad \int dp'_0\, G^I\,(p'_0,p_0)\,=\,\,\tau^2\,G^0(p_0)\,.\label{80}\end{equation}
Performing a 3D reduction as in subsection 2.3, we get again
\begin{equation}\phi\,=\,g^0\,\tau\,<\widetilde K^T>\,\phi\label{81}\end{equation}
with the definitions    
\begin{equation}\phi\,=\,\tau\,\sqrt{\omega}\int
dp_0\,\Phi(p_0)\label{82}\end{equation}
\begin{equation}\widetilde K^T\,=\,K(1-\widetilde
G^RK\,)^{-1},\qquad \widetilde G^R\,=\,G^0- G^I\label{82b}\end{equation}
and 
\begin{equation}<A>={1\over(-2i\pi)^{N-1}}\,{\tau^2
\sqrt{\omega}\over g^0}\int
dp'_0dp_0\,G^0(p'_0)A(p'_0,p_0)G^0(p_0)\,\beta\,{\tau^2
\sqrt{\omega}\over g^0}
\label{83}\end{equation} so that
\begin{equation} G^I=\,\,\,\,>\tau
g^0\!<\,.\label{84}\end{equation}
As in the two-fermion problem, we can choose to work in the
$\,\Lambda^+\!=\!1\,$ subspace instead of the $\,\tau^2\!=\!1\,$ subspace, by
simply replacing $\,\tau\,$ by $\,\Lambda^+.\,$ In this case also it is
possible to start the 3D reduction by taking the retarded part of the full
propagator at equal times \cite{G1,G2}. 

\subsection{Scattering.} 
As in the two-fermion case, we have $\,\widetilde T^{3D}\!=<T>.\,$ This implies
that $\,\,\widetilde T^{3D}\,$ is equal to the physical scattering amplitude
when all particles are on their positive-energy mass shells in their initial and
final states. In actual scattering experiments, however, the initial state
consists in two clusters only.\par
As shown already in \cite{G1}, the 3D equation remains valid in these cases.
The scattering of a bound state of fermions (23) by fermion 1 to a bound state
of fermions (12) plus fermion 3, for example, can be described by a part of the
full propagator $\,G:$
\begin{equation}G\,=\,G^0_3\,G_{12}\,T_{12,23}\,G_{23}\,G^0_1\,+\,\cdots
\label{s1}\end{equation}
where $\,T_{12,23}\,$ contains neither initial (23) nor final (12)
interaction, these interactions being included in $\,G_{23}\,$ and $\,G_{12}\,$
respectively. The full (23) propagator can be expanded as \cite{ST,15} 
\begin{equation}G_{23}(
P_{23},p'_{23},p_{23})\,=\,\Phi_{23}(\vec P_{23},p'_{23})\,{-i\over
P_{230}-E_{23}+i\epsilon}\,\overline\Phi_{23}(\vec P_{23},p_{23})\,+\,\cdots
\label{s2}\end{equation} 
where we isolated the pole at $\,P_{230}\!=\!E_{23}\!=\!\sqrt{\vec
P_{23}^2\!+\!P_{23}^2}\,$, defining $\,\overline \Phi_{23}$ as
$\,\Phi^+_{23}\,\beta_2\beta_3.\,$ The full (12) propagator can be expanded
similarly. The N-fermion 3D full propagator is, according to
(\ref{83}):    
\begin{equation}g\,=\,{1\over(-2i\pi)^{N-1}}\,\Lambda^+\,\int
dp'_0dp_0\,G(p'_0,p_0)\,\beta\,\Lambda^+.\label{s3}\end{equation}
Taking $\,N\!=\!3\,$ and putting (\ref{s1}) into (\ref{s3}), we see that 
the integration of $\,\overline \Phi_{23}(\vec P_{23},p_{23})\Lambda^+_{23}\,$ with
respect to $\,p_{230}\,$ will give $\,\overline \phi_{23}(\vec
P_{23},\vec p_{23})\,$ while in the integration with respect to $\,p_{10}\,$
we shall isolate the residue of the pole of $\,G^0_1\,$ at
$\,p_{10}\!=\!E_1\,$ (and similarly for the final state). This leads to
\begin{equation}g=-\phi_{12}{\Lambda^+_3\over
P_0-E_{12}-E_3+i\epsilon}\left[\overline
\Phi_{12}T_{12,23}\Phi_{23}\right]^{\delta\delta}_{31}{\beta_1\Lambda^+_1\over
P_0-E_{23}-E_1+i\epsilon}\phi^+_{23}+\cdots\label{s4}\end{equation}
where
\begin{equation}\left[\overline
\Phi_{12}T_{12,23}\Phi_{23}\right]^{\delta\delta}_{31}\equiv\int
dp'_0dp_0\,\delta(p'_{30}\!-\!E_3)T_{12,23}(p'_0,p_0)\delta(p_{10}\!-\!E_1).
\label{s5}\end{equation}
If we now compute $\,g\,$ at the 3D level, using the free propagator
$\,\Lambda^+g^0\,$ and the potential $\,\widetilde V,\,$ we get 
\begin{equation}g\,=\,\Lambda^+_3\,g_{12}\,\widetilde
T^{3D}_{12,23}\,g_{23}\,\Lambda^+_1\,+\,\cdots\label{s6}\end{equation}
while $\,g_{23}\,$ is obtained from $\,G_{23}\,$ through (\ref{s3}):
\begin{equation}g_{23}\,=\,{1\over2\pi}\,\phi_{23}\,{1\over
P_0-E_{23}-E_1+i\epsilon}\,\phi^+_{23}+\cdots\label{s7}\end{equation}
and similarly for $\,g_{12}.\,$ If we identify
(\ref{s4}) to (\ref{s6}) in momentum space, we get the following relation between
the physical scattering amplitudes computed at the 3D and at the 4D levels:
\begin{equation}
\phi_{12}^+\,\widetilde T^{3D}_{12,23}\,\phi_{23}\,=\,(-2i\pi)^2\,\left[\overline
\Phi_{12}\,T_{12,23}\,\Phi_{23}\right]^{\delta\delta}_{31}\,\beta_1\label{s8}\end{equation}
the momenta being such that 
$\,P_0\!=\!E'_{12}\!+\!E'_3\!=\!E_{23}\!+\!E_1.\,$ \par
Here, $\,\Phi_{23}\,$ depends on $\,\vec P_{23},p_{23}\,$ while
$\,\phi_{23}\,$ depends on $\,\vec P_{23},\vec p_{23}.\,$ We could compute
$\,\phi_{23}\,$ at $\,\vec P_{23}\!=\!0,\,$ then compute $\,\Phi_{23}\,$ from
it, boost $\,\Phi_{23}\,$ to $\,\vec P_{23}\!\neq\!0,\,$ and use it in
the right-hand side of (\ref{s8}), or we could try to compute directly
$\,\phi_{23}\,$ at $\,\vec P_{23}\!\neq\!0,\,$ and use it in the left-hand
side of (\ref{s8}). \par
We shall stop here these considerations about the
scattering, which is not the main subject of this work and would be better
studied on a specific example. We only wanted to show how the 3D equation
not only gives the bound state spectrum, but is also valid for the
scattering amplitudes.

\subsection{Reduction of the Dirac spinors.} Let us first consider a fermion in a
positive-energy potential:
\begin{equation}(p_0-h)\,\Psi=V\,\Psi,\qquad
V=\Lambda^+V\Lambda^+.\label{d1}\end{equation} Using the $\,\Lambda^+\,$
projector:
\begin{equation}(p_0-E\,)\,\Psi^+=V\,\Psi^+,\qquad
\Psi^+=\Lambda^+\Psi.\label{d2}\end{equation} We shall apply a virtual boost on
$\,\Psi^+:$
\begin{equation}\Psi^+\,=\,{1+(\gamma\cdot n)\,(\gamma\cdot
u\,)\over\sqrt{(u+n)^2}}\,\Psi^0\,,\qquad\Psi^0\,=\, {1+(\gamma\cdot
u\,)\,(\gamma\cdot n)\over\sqrt{(u+n)^2}}\,\Psi^+.\label{d3}\end{equation} with
\begin{equation}n\,=\,{1\over m}\,(E,\,\vec p\,),\qquad u\,=\,(1,\,\vec
0\,).\label{d4}\end{equation} In terms of $\,\Psi^0,\,$ equation (\ref{d2})
becomes
$$(P_0-E\,)\,\Psi^0\,=\,{1+(\gamma\cdot u\,)\,(\gamma\cdot
n)\over\sqrt{(u+n)^2}}\,\,V\,\,{1+(\gamma\cdot n)\,(\gamma\cdot
u\,)\over\sqrt{(u+n)^2}}\,\Psi^0$$
\begin{equation}=\,{1+\beta\over2}\,\sqrt{2\,m\over E+m}\,\,V\,\, {2\,E\over\sqrt
{2\,m\,(E+m)}}\,\Psi^0.\label{d5}\end{equation} Writing
\begin{equation}\Psi^0\,=\,\sqrt{m\over E}\,\left({\varphi\atop
0}\right)\label{d6}\end{equation} we get
\begin{equation}(p_0-E\,)\,\varphi\,=\,\sqrt{2\,E\over E+m}\,\,v\,
\,\sqrt{2\,E\over E+m}\,\,\varphi\label{d7}\end{equation}  where $\,v\,$ is the
large-large part of $\,V.\,$ Similarly, for f fermions and a positive-energy
integrating propagator-based reduction, we have
\begin{equation}(P_0-E\,)\,\varphi\,=\,\left[\,\prod_{i=1}^f\sqrt{2\,E_i\over
E_i+m}\,\,\right]\,\,v\,
\,\left[\,\prod_{i=1}^f\sqrt{2\,E_i\over
E_i+m}\,\,\right]\,\,\varphi.\label{d8}\end{equation}                

\section{Example: two and three fermions in QED.} Our aim in this section is only
to show how the integrating propagator-based 3D reduction method works for a
system of $\,N>2\,$ particles, since, as in any actual problem, specific
complications are encountered (choice of the gauge, estimation of the order of the
contributions for the exchange of zero-mass quanta, detection of the
cancellations...). In Coulomb's gauge the
relative energy dependence of the Bethe-Salpeter kernel lies in the transverse
part, the contributions of which are typically smaller by a factor
$\,\alpha^2.\,$  This makes the calculation of the first-order contributions (up
to $\,\alpha^4$) more easy, but does not provide a good illustration of our 3D
reduction method, the three-body contributions to the 3D potential being
then of order
$\,>\alpha^4\,$ and competing with a lot of other effects. We shall therefore
work in Feynman's gauge until subsection 4.4.  

\subsection{Two fermions.} For two fermions of charges $\,Z_1,Z_2\,$ the
Bethe-Salpeter kernel is
$$K(p',p)\,=\,{2i\,Z_1Z_2\over(2\pi)^3}\,\,{1\over k^2+i\epsilon}\,
(\gamma_1\!\cdot\gamma_2)$$
\begin{equation}+\left[{2i\,Z_1Z_2\over(2\pi)^3}\right]^2\int dk\,\,{1\over
k'^2+i\epsilon}\,\,\,{1\over
k^2+i\epsilon}\,\,\gamma_1^{\mu'}\gamma_2^{\mu}\,\,G^0_1(q_1)\,G^0_2(q_2^c)\,\,
\gamma_{1\mu}\gamma_{2\mu'}\,\,+\cdots\label{85}\end{equation}  We wrote explicitly the ladder term
$\,K^L\,$ and the first crossed term $\,K^C.\,$ The meaning
of the variables can be read on figure 1, which represents\break $\,<\!K^L\!>,
<\!K^C\!>\,$ and $\,<\!K^L\widetilde G^RK^L\!>\,=\,<K^L
G^0K^L>\!-\!<\!K^L\!>\!g^0\!<\!K^L\!>.\,$   
The results of subsections 2.2 to
2.4 will be used, and specialized to the kernel (\ref{85}).     If we choose to
perform the 3D reduction in the
$\,\Lambda^{++}\!=\!1\,$ subspace, we get
\begin{equation}<K^L>(\vec p\,',\vec
p\,)\,=\,{2i\,Z_1Z_2\over(2\pi)^3}\,\Lambda^{++}(\vec
p\,')\,(1-\vec\alpha_1\!\cdot\!\vec\alpha_2)\,\Lambda^{++}(\vec p\,)\,I\,(\vec
p\,',\vec p\,)\label{86}\end{equation}
$$I\,(\vec p\,',\vec
p\,)\,=\,{(P_0\!-\!E'_1\!-\!E'_2\,)(P_0\!-\!E_1\!-\!E_2\,)\over-2i\pi}
\int dp'_0dp_0\,\,{1\over p'_{10}\!-\!E'_1+i\epsilon}\,\,{1\over
p'_{20}\!-\!E'_2+i\epsilon}$$
$${1\over k^2+i\epsilon}\,\,\,{1\over p_{10}\!-\!E_1+i\epsilon}\,\,{1\over
p_{20}\!-\!E_2+i\epsilon}$$
\begin{equation}=\,{-2i\pi\over (E'_1\!-\!E_1)^2-\vec
k\,^2+i\epsilon}\,\,\left[\,1+{1\over2\,\vert\vec
k\vert}\,R\,\right]\label{87}\end{equation}
\begin{equation}R\,=\,{(E'_1\!-\!E_1\!-\!\vert\vec
k\vert\,)\,(P_0\!-\!E_1\!-\!E_2)\over P_0\!-\!E'_1\!-\!E_2\!-\!\vert\vec
k\vert+i\epsilon}\,-\,{(E'_1\!-\!E_1\!+\!\vert\vec
k\vert\,)\,(P_0\!-\!E'_1\!-\!E'_2)\over P_0\!-\!E_1\!-\!E'_2\!-\!\vert\vec
k\vert+i\epsilon}\label{88}\end{equation} where the first part of $\,I\,$
corresponds to $\,<K^L>^{\delta\delta}\,$ and contributes in $\,\alpha^2\,$ to
the energy. The two terms of $\,R\,$ correspond to $\,<K^L>^{\delta R}\,$ and
$\,<K^L>^{R\delta}\,$ respectively. At lowest order
\begin{equation}R\,\approx\,(P_0\!-\!E_1\!-\!E_2)\,+\,(P_0\!-\!E'_1\!-\!E'_2).
\label{89}\end{equation} By sheer power counting (pc), $\,R\,$ should contribute
in $\,\alpha^3\,$ ($\,\vert\vec k\vert^{-1}\,$ brings a factor $\,\alpha^{-1}\,$
and the difference of the energies a factor $\,\alpha^2,\,$ to be combined with
the
$\,\alpha^2$ of Coulomb's potential). However,  this dominant contribution, if
isolated, is in $\,\vert\vec k\vert^{-3}\,$ and makes the future integrations
with respect to $\,\vec k\,$ diverge at origin. A more careful estimation is thus
necessary, and shows that
$\,R\,$ contributes in fact in $\,\alpha^3log(\alpha).\,$ This divergence of the
dominant contribution is a consequence of the zero mass of the photon. From now
on, we shall estimate only the (pc) order of the contributions, keeping in mind
that the real order could be lower.\par The expression of
$\,I\,$ is apparently asymmetric for the permutation of the fermions, but it can
be shown that it is in fact equal to the expression obtained after permutation or
the fermions, on by directly closing the integration paths counterclockwise. The
differences between
$\,[\,-\vec k\,^2]^{-1},\,$
$\,[\,(E'_1\!-\!E_1)^2\!-\!\vec k\,^2]^{-1}\,$ and
$\,[\,(E'_2\!-\!E_2)^2\!-\!\vec k\,^2]^{-1}\,$ are in
$\,\alpha^4\,$ (pc) (these three denominators are equal in the equal mass case if
we work in the two-fermion rest frame).\par A third way of computing $\,I\,$
consists in writing the photon's pole as
\begin{equation}{1\over k^2+i\epsilon}\,=\,{-1\over2\,\vert\vec
k\vert}\,\left[\,{1\over p'_0-p_0+\vert\vec k\vert-i\epsilon}\,\,-\,\,{1\over
p'_0-p_0-\vert\vec k\vert+i\epsilon}\right]\label{90}\end{equation}  and closing
for each part of (\ref{90}) the integration path in order to leave the photon's
pole outside, with the result 
\begin{equation}I\,(\vec p\,',\vec p\,)\,=\,{2i\pi\over2\,\vert\vec
k\vert}\,\left[\,{1\over E'_1+E_2-P_0+\vert\vec k\vert-i\epsilon}\,\,+\,\,{1\over
E_1+E'_2-P_0+\vert\vec k\vert-i\epsilon}\right]\label{91}\end{equation} which is
the expression obtained in time-ordered perturbation theory \cite{25},  and again
a rearrangement of (\ref{87}).\par If we include now the four-vertex graphs, we
must compute
\begin{equation}\widetilde V\,\approx\,\,<K^L\!>\!+\!<K^C\!>+<K^L\,
G^0K^L\!>-<K^L\!> g^0\!<K^L\!>,\label{92}\end{equation}  of which we already have
$\,<K^L\!>\!.\,$ If we choose to split $\,G^0\,$ into
$\,G^{\delta}\!+\!G^R,\,$ we get, keeping only the terms with zero or one
$\,G^R,\,$
$$\widetilde V\approx\,<K^L>^{\delta\delta}\,+\,<K^C>^{\delta\delta}\,+\,
<K^LG^RK^L>^{\delta\delta}$$
\begin{equation} +\,\,\big[1-<K^L>^{\delta\delta}
g^0\big]<K^L>^{R\delta}+<K^L>^{\delta R}\big[1-
g^0<K^L>^{\delta\delta}\big].\label{93}\end{equation} The second line will not
contribute to the first-order energy shift of a perturbation calculation starting
with $\,<K^L\!>^{\delta\delta}.\,$ If we start with a Coulomb potential, then
$\,<K^L\!>^{\delta\delta}\,$  will provide perturbations in $\,\alpha^4\,$ (pc)
and the second line perturbations in $\,\alpha^5\,$ (pc). The two last terms of
the first line are in
$\,\alpha^3\,$ (pc), but their higher-order contributions cancel mutually,
leaving contributions in $\,\alpha^4\,$ (pc).    

\subsection{Three fermions.}
 For the three-fermion problem,
we can write
\begin{equation}\phi\,=\,g^0\,\widetilde V\,\phi,\qquad
\widetilde V=\,<K>+<KG^0K>-<K>g^0<K>+\cdots\label{94}\end{equation} with the
definitions
$$<A>\,=\,\,\,<<G^0A\,G^0>>$$
\begin{equation}=\,\,{-1\over4\pi^2}\,\Lambda^{+++}(g^0)^{-1}\int
dp'_0dp_0\,G^0(p'_0)A(p'_0,p_0)G^0(p_0)\,\beta_1\beta_2\beta_3\,\Lambda^{+++}
\,(g^0)^{-1}
\label{95}\end{equation}
\begin{equation}g^0\,=\,{1\over P_0-E_1-E_2-E_3+i\epsilon}\,,\qquad
G^0=G^0_1\,G^0_2\,G^0_3\label{96}\end{equation}
\begin{equation}K\,=\,K_{12}\,(G^0_3)^{-1}\,+\,
K_{23}\,(G^0_1)^{-1}\,+\,K_{31}\,(G^0_2)^{-1}\,+\, K_{123}\label{97}\end{equation}
\begin{equation}dp_0\,=\,\delta(P_0-p_{10}-p_{20}-p_{30})\,dp_{10}
\,dp_{20}\,dp_{30}.\label{98}\end{equation}   We introduced the notation
$\,<<>>\,$ in order to exhibit the $\,G^0_i\,$ operators below.\,   The
$\,K_{ij}\,$ will be given by (\ref{85}). We shall here neglect the three-fermion
irreducible kernel which begins with 6-vertex interactions only. The two- and
4-vertex diagrams contributing to $\,\widetilde V\,$ are given in figure 2 (we
draw only 
$\,L_3, C_3,B_3\,$ and $\,L_{31},\,$ as the other ones can be obtained by
two-fermion permutations and/or three-fermion circular permutations). 
The computation of these
diagrams is tedious but straightforward. Each external fermion line represents a
propagator
$\,(p_{i0}\!-\!E_i+i\epsilon)^{-1}\,$ and each internal fermion line a propagator
$\,(p_{i0}\!-\!h_i+i\epsilon h_i)^{-1}.\,$ We could integrate clockwise with
respect to the energies on the lines 1 and 3 (for the diagrams drawn in the
figure), replacing $\,p_{20}\,$ by
$\,P_0\!-\!p_{10}\!-\!p_{30}.\,$ The leading contributions will come from the
 four residues of the positive-energy fermion poles on lines 1 and 3, while
higher-order contributions will appear when one or several of these poles are
replaced by a photon pole or by an internal fermion 2 pole (in $\,C_3\,$ the
residues on the internal 1 and 2 lines are both leading terms, if we close the
integral on $\,p_{10}\,$ clockwise, but they cancel mutually at leading order).
\par For the sum of the (12)+3 unconnected graphs, the residue of the pole in
$\,(p_{30}\!-\!E_3+i\epsilon)^{-1}\,$ will give the (12) potential, with the (12)
energy $\,P_{120}\,$ replaced by
$\,P_0\!-\!E_3\,$ (if we close the $\,p_{30}\,$ integration path clockwise).  The
sum of the residues of the propagator of the spectator fermion in the unconnected
graphs will thus give the sum of the three two-fermion potentials:
\begin{equation}\widetilde V_u\,=\,<\widetilde K^T_{12}>_{P_{120}=P_0-E_3}
+<\widetilde K^T_{23}>_{P_{230}=P_0-E_1}+\!<\widetilde
K^T_{31}>_{P_{310}=P_0-E_2}\label{aa}.\end{equation}   In graphs like $\,C_3\,$
and $\,B_3,\,$ however, the internal fermion
propagators will also provide poles inside the
$\,p_{30}\,$ integration path, unless we neglect the negative energy parts. These
contributions (which can be neglected at order
$\,\alpha^4\,$)  apparently violate cluster separability, even if we do not
choose the reference frame as being the global rest frame and left it
unspecified. They do not vanish indeed when all interactions between particle 3
and the (12) cluster are "switched off". However, such a "switching off" is of
course not compatible with the existence of a global bound state, and we have
shown above that our 3D equation gives anyway the correct cluster-particle
scattering amplitudes.\par The connected graph
$\,L_{13}\,$ is more specific of a three-body problem. It is given by
$$L_{13}\,=\,\,<K^L_{23}\,G^0_2\,K^L_{12}>\,-\,<K^L_{23}\,
(G^0_1)^{-1}>g^0<\,K^L_{12}\,(G^0_3)^{-1}>$$
$$=\,\,<<G^0_1\,G^0_2\,G^0_3\,K^L_{23}\,G^0_2\,K^L_{12}\,G^0_1\,G^0_2\,G^0_3\,>>$$
\begin{equation}-\,\,<<G^0_1\,G^0_2\,G^0_3\,K^L_{23}\,G^0_2\,
G^0_3>>g^0<<G^0_1\,G^0_2\,K^L_{12}\,G^0_1\,G^0_2\,G^0_3\,>>.\label{99}
\end{equation}         
Each $\,G^0_1\,$ and $\,G^0_3\,$ can be splitted  into
\begin{equation}G^0_i\,=\,G^{\delta}_i+G^R_i\,,\qquad 
G^{\delta}_i\,=\,-2i\pi\,\Lambda^+_i\,\delta\,(\,p_{i0}\!-\!E_i\,)\,\beta_i\,.
\label{100}\end{equation} Let us consider the initial $\,G^0_3\,$ and the final
$\,G^0_1\,$ in
$\,L_{13}.\,$ There is a mutual cancellation between the two terms of
$\,L_{13}\,$ when these propagators are both replaced by their $\,\delta\,$ part
(this suppression of the leading contribution is indeed at the basis of our
reduction method). If we keep only the one-$G^R\,$ contributions to
$\,L_{13},\,$ we get
$$L_{13}\,\approx\,\,\,<<G^{\delta}_1\,G^0_2\,G^{\delta}_3\,K^L_{23}
\,G^0_2\,K^L_{12}
\,G^{\delta}_1\,G^0_2\,G^R_3\,>>$$
$$\,+\,<<G^R_1\,G^0_2\,G^{\delta}_3\,K^L_{23}
\,G^0_2\,K^L_{12}
\,G^{\delta}_1\,G^0_2\,G^{\delta}_3\,>>$$
\begin{equation}+\,\,<K^L_{23}>^{\delta\delta}_{33}
\,\,g^0<K^L_{12}>^{R\delta}_{11}
\,\,+\,\,<K^L_{23}>^{\delta R}_{33}
\,\,g^0<K^L_{12}>^{\delta\delta}_{11}\label{101}\end{equation}  where the indexes
11 or 33 indicates which $\,G^0_i\,$ is replaced by
$\,G^{\delta}_i\,$ or $\,G^R_i.\,$ The two first terms of (\ref{101}) will
contribute in $\,\alpha^4\,$ (pc). The two other terms will contribute in
$\,\alpha^3\,$ (pc) and the factors inside them will be given, up to permutations,
by one of the three terms of (\ref{86}-\ref{87}), with the spectator fermion on
its positive-energy mass shell. As in the two-fermion problem, however, the
contributions of these terms will be of higher-order in a perturbation
calculation: changes of indexes $\,<K_{ij}>_{ii}\,\to\, <K_{ij}>_{jj}\,$
correspond to changes in $\,\alpha^4\,$ (pc) in (\ref{101}). If we isolate a
leading contribution in $\,\alpha^3\,$ (pc) we can built combinations like
\begin{equation}\left[1-(<K^L_{23}>^{\delta\delta}
+<K^L_{31}>^{\delta\delta}+<K^L_{12}>^{\delta\delta})\,
g^0\,\right]\,<K^L_{12}>^{R\delta}\label{102}\end{equation} coming from
$\,L_3,B_3,L_{13}\,$ and $\,L_{23}.$\par If we perform the energy integrations of
(\ref{101}), we found, for the first and the last terms
$$L_{13}^{(1+4)}(\vec p\,'\!\!,\vec p\,)
\,=\,{-4\over(2\,\pi)^6}\,Z_1Z_2^2Z_3\,\Lambda^{+++}(\vec p\,')
\,(1-\vec\alpha_2\!\cdot\!\vec\alpha_3)$$
\begin{equation}\,\Lambda_2^+(\vec q_2)\,(1-\vec\alpha_1
\!\cdot\!\vec\alpha_2)\,\Lambda^{+++}(\vec
p\,)\,I_{13}^{(1+4)}(\vec p\,'\!\!,\vec p\,),\label{101a}\end{equation}
$$I_{13}^{(1+4)}(\vec p\,'\!\!,\vec p\,)=-4\pi^2\,{1\over2\,\vert\vec
k\,'_{23}\vert}\,\,\,{1\over P_0\!-\!E'_1\!-\!E_2(\vec
q_2)\!-\!E'_3\!-\!\vert\vec
k\,'_{23}\vert}\,\,\,{1\over(E'_1\!-\!E_1)^2\!-\!\vert\vec k_{12}\vert^2}\,\,\,$$
\begin{equation}{1\over E'_3\!-\!E_3\!+\!\vert\vec k\,'_{23}\vert}\,\,\,\left[{
P_0\!-\!E_1\!-\!E_2\!-\!E_3\over P_0\!-\!E_1\!-\!E_2\!-\!E'_3\!-\!\vert\vec
k\,'_{23}\vert}\,\,\,+\,\,\,1\,\right]\label{101b}\end{equation} where we
neglected the $\,\Lambda_2^-(\vec q_2)\,$ part of the internal $\,G^0_2(q_2)\,$
propagator and its residue. The $\,L_{13}^{(2+3)}\,$ part of $\,L_{13}\,$ in
obtained by the permutations
$$\vert\vec k\,'_{23}\vert\leftrightarrow\vert\vec k_{12}\vert,\quad
E_1\leftrightarrow E'_3,\quad E'_1\leftrightarrow E_3,
\quad E_2\leftrightarrow E'_2.$$

\subsection{Gross' spectator model.} 
When the three-fermion irreducible
Bethe-Salpeter kernel is neglected, the global propagator becomes a sequence of
two-fermion interactions, for which the third fermion is a spectator. In Gross'
spectator model \cite{26,27}, the spectator fermion is put on its positive-energy
mass shell (we could say that we keep only the $\,G^{\delta}_i\,$ part of the
corresponding propagator $\,G^0_i\,$). When a (12) interaction, for example, is
followed by a (23) interaction (first part of figure 3), the fermions 1 and 3
between them are put on their positive-energy mass shell. The neglected
$\,G^R_i\,$ parts and the neglected three-body irreducible parts of the kernel
are then treated as higher-order corrections. The result is a set of 3D equations
for partial transition operators, which can not be reduced to a single equation
as the initial and final "spectator" fermions are not the same in all
terms.\par    In the integrating propagator based reduction, all terms of the
transition operator are treated in the same way. We expect of course that the
most important contributions to the integrals in the relative energies will be
the residues of these propagator's poles which put the spectator fermions on their
positive-energy mass shell. The 3D potential $\,\widetilde V\,$  differs from the
integrated transition operator $\,<T>\,$ by the presence of the  counter-terms in
$\,-\!>\!g^0\!\!<\,,\,$ as shown in figure 3. If we put the spectator
fermions on their positive-energy mass shell in these terms too, then the two
diagrams of figure 3 cancel mutually, and, finally,
there remains only the unconnected contributions (\ref{aa}) to the
potential. This is in fact the kind of potential we proposed in ref. \cite{20}
(in which there is also a comparison with Gross' spectator model). The 
$\,G^R_i\,$ parts of the free propagators of the spectator fermions and the
irreducible three-body terms provide corrections to it.   

\subsection{Coulomb gauge.} In Coulomb's gauge, we would make the replacement
\begin{equation}{1\over k^2+i\epsilon}\,
\beta_1\beta_2\,(\gamma_1\!\cdot\gamma_2)\,\rightarrow\, {1\over-\vec
k\,^2}\,\,-\,\,{1\over k^2+i\epsilon}\,
(\vec\alpha_1\!\cdot\vec\alpha_2)^T\label{102a}\end{equation} 
\begin{equation}(\vec\alpha_1\!\cdot\vec\alpha_2)^T\,=\,
(\vec\alpha_1\!\cdot\vec\alpha_2)\,-\,{(\vec\alpha_1\!\cdot\vec k)\,
(\vec\alpha_2\!\cdot\vec k\,)\over \vec k\,^2}.\label{103}\end{equation} In
(\ref{102a}), the relative-energy dependence lies in the second (transverse) term.
The contribution of this term to the energy begins at $\,\alpha^4,\,$ as it
connects the positive-energy components with the negative-energy components. At
this order, we can replace $\,k^2\,$ by $\,-\vec k\,^2.\,$ In the two-fermion
problem, we get
\begin{equation}I\,(\vec p\,',\vec p\,)\,=\,{2i\pi\over \vec
k\,^2}\label{104}\end{equation}
\begin{equation}\widetilde V\,(\vec p\,',\vec
p\,)\,=\,{-1\over2\pi^2}\,\,{Z_1Z_2\over \vec k\,^2}\,\Lambda^{++}(\vec
p\,')\,(1-(\vec\alpha_1\!\cdot\!\vec\alpha_2)^T\,)\,\Lambda^{++}(\vec
p\,).\label{105}\end{equation} In the three-fermion problem, up to order
$\,\alpha^4,\,$ we simply must add the three Coulomb and the three Breit
potentials:
$$\widetilde V\,(\vec p\,',\vec p\,)\,=\, \widetilde V_{12}\,(\vec p\,'_{12},\vec
p_{12}\,)\,\delta^3(\vec p\,'_3\!-\!\vec p_3)$$
\begin{equation}\,+\,\,\,\widetilde V_{23}\,(\vec p\,'_{23},\vec
p_{23}\,)\,\delta^3(\vec p\,'_1\!-\!\vec p_1)\,+\,\widetilde V_{31}\,(\vec
p\,'_{31},\vec p_{31}\,)\,\delta^3(\vec p\,'_2\!-\!\vec
p_2).\label{106}\end{equation} The contributions which are specific to the
three-body problem will thus appear beyond $\,\alpha^4.$        

\section{Conclusions} Our integrating propagator-based reduction of the
Bethe-Salpeter equation provides a quite straightforward way of writing the
different terms of the 3D potential for any N-particle bound state equation. The
continuum-dissolution problem is automatically avoided. Our reduction provides a
link between the kernel-based reductions (especially that of Phillips and Wallace
\cite{19}) and the equal-time retarded full propagator method of Logunov and
Tavkhelidze
\cite{4} and Kvinikhidze and Stoyanov \cite{G1,G2}. Although first developed
for the two-fermion bound state problem, it has been naturally generalized to the
N-particle bound state problem and is also valid for the 2-body and
particle-cluster scattering problems (even if one may prefer another approach in
this case). The 3D equation provides a (in principle) indefinitely improvable
approach of the exact values of the measurable quantities, despite the fact that
the individual terms of the series giving the 3D potential are neither Lorentz
covariant nor all cluster separable.

\end{document}